\documentclass[a4paper,11pt]{iopart}
\usepackage{graphicx,sidecap}
\usepackage{bm}
\usepackage{amssymb}
\usepackage{cite}
\usepackage{amsfonts}
\usepackage{mathrsfs}
\usepackage{xcolor}
\usepackage{hyperref}
\usepackage{enumerate}
\usepackage{verbatim}
\usepackage[margin=30pt, bf, font=footnotesize, center, justification=justified]{caption}[2004/07/16]
\usepackage{subfigure}
\usepackage{multirow}
\usepackage{soul}

\hypersetup{colorlinks,bookmarksopen,bookmarksnumbered,
citecolor=red,
linkcolor=blue,
pdfstartview=green,
urlcolor=teal}

\def\d{\mathrm{d}}
\def\Li{\mathrm{Li}}
\def\kB{k_\mathrm{B}}
\def\diag{\mathrm{diag}}

\definecolor{myred}{rgb}{0.9, 0.15, 0.24}

\begin{document}

\title[Long-range interacting chain with long-range
conservative noise]{Thermal transport in long-range interacting
  harmonic chains perturbed by long-range conservative noise}

\author{Francesco Andreucci$^{1}$, Stefano Lepri$^{2,3}$, Carlos
  Mej\'ia-Monasterio$^{4,5}$,
Stefano Ruffo$^{1,2}$}

\address{$^{1}$ SISSA, Via Bonomea 265,
  34136 Trieste, Italy}
\address{$^{2}$ Istituto dei Sistemi Complessi, Consiglio Nazionale
delle Ricerche, via Madonna del Piano 10, 50019 Sesto Fiorentino, Italy}
\address{$^{3}$ Istituto Nazionale di Fisica Nucleare, Sezione di
    Firenze, via G. Sansone 1, 50019 Sesto Fiorentino, Italy}
\address{$^{4}$ School of Agricultural, Food and Biosystems Engineering,
  Technical University of Madrid, Av. Puerta de Hierro 2, 28040
  Madrid, Spain}
\address{$^{5}$ Grupo Interdisciplinar de Sistemas Complejos (GISC), Spain}

\eads{\mailto{\mailto{fandreuc@sissa.it}, carlos.mejia@upm.es},
  \mailto{stefano.lepri@isc.cnr.it}, \mailto{ruffo@sissa.it}}

\begin{indented}
\item[]\today
\end{indented}

\begin{abstract} 
  We study non-equilibrium properties of 
  a chain of $N$ oscillators with both long-ranged harmonic interactions and long-range conservative noise that exchange momenta of particle pairs. 
  We derive exact expressions for the (deterministic) energy-current auto-correlation at equilibrium, based on the kinetic approximation of the normal mode dynamics. In all cases the decay is algebraic in the thermodynamic limit. We distinguish 
four distinct regimes of correlation decay  depending on the exponents controlling the range of deterministic and stochastic 
interactions.  
  Surprisingly, we find that long-range noise breaks down the long-range correlations characteristic of low dimensional  models, suggesting a normal regime in which  heat transport becomes diffusive. For finite systems, we do also derive
  exact expressions for the finite-size corrections to the algebraic decay of the correlation.  
  In certain regimes, these corrections are considerably large, rendering hard the estimation of transport properties from numerical data for the finite chains. Our results are tested against numerical simulations, performed with an efficient algorithm.
\end{abstract}
\vspace{2pc}

\section{Introduction}
\label{sec:intro}

Heat and mass flow through  a medium is a familiar thermodynamic phenomenon
relevant for both basic physics and technology. 
From the point of view of statistical physics,  the microscopic foundations
of its macroscopic laws 
have challenged researchers for decades. In this context,
the possibility
of anomalous  energy transport and violations of Fourier's law in low-dimensional non-linear systems has been thoroughly investigated \cite{lepri2003,dhar_heat_2008,benenti2020anomalous}. This problem is of interest
for nanoscale heat transfer and thermal management, see the recent review \cite{benenti2023non}.
Anomalous heat conduction in low-dimensional 
many-particle systems has been studied by simulations in 
many works.
In one dimension the main finding is that,  the long-time, nonintegrable, 
tail of the current correlation decays at large times as $t^{-\beta}$, 
with $\beta<1$.  Typically this entail that, for a finite system of length $L$
where heat carriers have a finite propagation velocity,the heat flux scales as $L^{-\beta}$ as well.  The main finding is that $\beta$
is largely independent on the microscopic details, being solely determined  
by the space dimensionality and the coupling among fluctuations of the 
conserved quantities.    
Within the nonlinear fluctuating hydrodynamics approach it has been indeed 
shown that of energy current correlations and the dynamical 
scaling of correlations in one-dimension, are universal and 
belong (generically) to the class of the famous Kardar-Parisi-Zhang
equation \cite{Spohn2014}. Some experimental evidence of 
superdiffusive heat transport in  carbon nanotubes \cite{Chang08}
and atomic chains \cite{yang2021observation} have been reported.

All the above applies to short-range forces (e.g. nearest-neighbor couplings
on the lattice).  One may wonder  about
the effect of long-range interactions, i.e.  
the case in which the interparticle potential decays at large 
distances $r$ as $ r^{-d-\sigma}$, in dimension $d$. \cite{campa_statistical_2009,Campa2014}.
The study of such forces is well developed in equilibrium statistical mechanics, starting from the seminal works by
 F.J. Dyson \cite{dyson1969existence}, D. Thouless  \cite{thouless1969long} and others. 

Out of equilibrium the problem is even more difficult and intriguing. Actually, 
for interactions decaying sufficiently slowly 
with distance, fluctuations may propagate with infinite velocities, yielding qualitative differences with respect to the short-ranged case \cite{Metivier2014}.
As far as transport and hydrodynamics are concerned, non-local effective equations are expected to arise naturally by the non-local 
nature of couplings \cite{Schukert2020}. This has also effects on energy transport for open systems interacting with external reservoirs and, more 
generally, on the way in which the long-range terms couple the system with 
external reservoirs.

Besides the theoretical motivations, there are also experimental systems
where those effects may be relevant, notably trapped ion chains, dipolar condensates
etc.  both classical and quantum \cite{defenu_long-range_2021}.
As a concrete experimental instance, we mention 
trapped ion chains, where ions are confined in periodic arrays and interact with thermal reservoirs  \cite{bermudez2013controlling,ramm2014energy}.
On a macroscale, effective long-range forces arise also in 
chains of coupled magnets \cite{moleron2019nonlinear} where nonlinear 
effects may be very relevant.

Several studies of low-dimensional, long-range interacting 
models appeared in the more recent
literature, notably for chains of coupled rotors \cite{olivares_role_2016}  and oscillators \cite{bagchi_energy_2017,iubini_heat_2018,
wang_thermal_2020,bagchi_heat_2021} under thermal gradients.
There is numerical evidence that 
non-Fourier transport occurs, albeit with characteristic exponents
depending on the range of the forces.
At equilibrium, correlations decay in a nontrivial way depending on the 
range exponent, suggesting that the hydrodynamic description may be 
nonstandard  \cite{di_cintio_equilibrium_2019,iubini2022hydrodynamics}.
Loosely speaking, on the hydrodynamic scales,  the energy carriers propagate
effectively as a L\'evy flight and fluctuations follow a fractional diffusion equation . For a finite system of length $L$, this entails a non-Fourier heat transport with a superdiffusive  scaling of the energy flux with
$L$. Moreover, an intriguing feature is that models having the same coupling $r^{-1-\sigma}$ may belong to different dynamical universality classes, having
different hydrodynamics (see the discussion in \cite{iubini2022hydrodynamics}).
Another novel feature of long-range forces 
is that they may yield phase 
transitions even in low dimensions. Thus their effect on transport can be studied even  
in a one-dimensional setup \cite{iubini2022hydrodynamics}.

It is noteworthy that also the simpler case of \textit{harmonic} long-range 
forces is far from trivial. Indeed, the case of lattices with mean-field \cite{defaveri_heat_2022,andreucci_classical_2022} and power-law 
decaying coupling  
\cite{andreucci_nonequilibrium_2023}
have been analytically considered, and several intriguing features have 
been demonstrated.  For instance, for strong long-ranged forces , the energy flux  shows 
anomalous  scaling  with the size and the 
plane-wave transmission spectrum acquires a self-similar structure \cite{andreucci_nonequilibrium_2023} .

Coming back to the general case,  one common difficulty is the need to deal with 
anharmonic forces. Although 
molecular dynamics simulations are in principle straightforward, the data are often
plagued by finite-size and time effects that often hinder conclusive comparisons
with the theories. An alternative approach rests on the 
stochastic modeling of the interaction on a mesoscopic level. This leads to 
a sort of hybrid dynamical system, where the deterministic nonlinearity is 
replaced by an effective stochastic interactions, under the basic requirements 
that the conservation laws of energy, momentum and density should be 
preserved. In the simplest setup, the deterministic dynamics is linear, 
while the random one provides ergodicity and a mechanism for energy diffusion. 
This is often referred to as \textit{conservative noise }dynamics: in its simplest versions
it entails random exchange of momenta between particles or a random 
reshuffling of a subset of particles,  like in the multi-particle-collision 
protocol 
see e.g.\cite{basile2006,Malevanets1999,Benenti2014,DiCintio2015}. 
This allows for very efficient simulations and,  in some 
simple  
cases for exact solutions.   
For the class of oscillator chains, this class of random dynamical systems
is amenable of mathematically  rigorous analysis
\cite{basile2006,basile2009,bernardin2012harmonic,basile_thermal_2015}.
Indeed, large-scale hydrodynamics equations can be demonstrated 
and  phonon Boltzmann equation can be derived, yielding relatively
simple linear collision operators  \cite{basile2010energy,lukkarinen2016harmonic}.
Moreover,  most of nonequilibrium steady-state
properties  can be computed exactly 
and were shown to reproduce many features of deterministic nonlinear lattices \cite{lepri2009,lepri2010,delfini2010,kundu2019} . The effect of conservative noise on nonlinear
oscillator chains has also been considered 
\cite{Iacobucci2010,bernardin2014anomalous,lepri2020}.

In the present work we consider a one-dimensional chain of 
harmonic oscillators with long-range couplings, decaying as an inverse 
power of the distance between sites.  On top, we have 
a conservative noise  that exchange 
momenta  of a pair of oscillators, indexed, say, by $(n,m)$ . 
The case of  long-range forces and nearest-neighbor
exchanges  ($m=n\pm 1$) has been studied 
in \cite{tamaki2020} 
(see also \cite{suda2022superballistic} for a mathematical analysis
of its hydrodynamic limit).
Here, we generalize to the case in which the exchange occurs between 
sites $n,m$ at \textit{arbitrary distances},  chosen with a probability decaying as a power
$|n-m|^{-\alpha}$ of their distance. This choice should mimic the effect of 
\textit{long-range anharmonic} forces. We mostly focus on the decay of the energy 
current autocorrelation that contains the most relevant information on transport
coefficients (the thermal conductivity here) and their anomalous behavior.
We study the dependence of the autocorrelation on the lattice size, and
we show that finite-size effects are of major importance. 
We follow an approach based on the kinetic approximation, proposed in \cite{lepri2023}.
It allows to estimate the characteristic relaxation rates of Fourier modes
quite easily. The information about the 
scaling of the rates with the wavenumber can be used to obtain analytical approximations
of the auto-correlations and to discuss their dependence on the exponents 
of the interactions.   We are thus able to reconstruct the phase diagram
for the decay of the correlations and to estimate the size dependence
of the conductivity in the various regimes. 

\section{Long-range interacting chain with long-range conservative noise}
\label{sec:model}

We  consider a  homogeneous one-dimensional  chain of  $N$ interacting
harmonic oscillators. 
The displacement from the equilibrium position and momentum of the $i$-th particle
are  denoted  as  $q_i$  and  $p_i$  respectively ($i=1,\ldots, N)$.   Without  loss  of
generality we set the particles' mass  to $m=1$ and spring constant to
$k=1$. The oscillators interact through a long-range potential that
decays as a power-law $r^{-\delta}$ of the distance $r$ between oscillators.
The dynamics of the system is determined by the Hamiltonian
\begin{equation} \label{H}
  H(\mathbf{q},\mathbf{p}) = \sum_{i=1}^N \frac{p_{i}^{2}}{2} +
  \frac{1 }{N_\delta} \sum_{i=1}^N \sum_{r=1}^{N/2} \frac{(q_{i+r} -
    q_i  )^2}{2 r^\delta} \ ,
\end{equation}
where 
 $N_\delta = \sum_{r=1}^N1/r^{\delta}$ is the
so-called Kac factor that ensures extensivity of the Hamiltonian.
Periodic boundary conditions are assumed so that $q_{i+N} = q_i$  and
$p_{i+N} = p_i$.   The dynamics
posses three conserved quantities: the total energy, the total
momentum, and the total stretch.

The deterministic  dynamics given  by \eref{H} is perturbed  by a
conservative noise  defined as follows:  with rate $\gamma$ a  pair of
oscillators  $n$  and $m$,  not  necessarily  nearest neighbours,  are
randomly chosen with probability $\mathcal{W}_{n,m}$. Then the momenta
of these two oscillators are exchanged, as if they experienced a front 
collision
\begin{equation}
    (p_{n},p_{m})\rightarrow  (p_{m},p_{n})\equiv(p_{n}^{'},p_{m}^{'}),
\end{equation}
where the prime denotes a quantity immediately after the event.
This  noise   was  termed  conservative   since , by
construction, it preserves the conserved quantities of \eref{H}.
It has been thoroughly 
studied mostly  for local 
random    collisions,  occurring  either between nearest-neighbours
$\mathcal{W}_{n,m} =  \delta_{|n-m|,1}$ or triplets of particles
~\cite{basile_thermal_2015}. 

Here we consider a long-range version of the
conservative noise for which the probability $\mathcal{W}_{n,m}$
decays as a power-law of their distance
\begin{equation} \label{Wprob}
  \mathcal{W}_\alpha(r) = \frac{1/r^\alpha}{\sum_{k=1}^{N/2} 1/r^\alpha} \ ,
\end{equation}
where $r = |n-m|$. The effective rate at which a pair of particles $j$
and $j+r$ exchanges their momenta is $\gamma \mathcal{W}_\alpha(r)$.

These dynamics include different model systems, some of which has been
studied intensively in the past. In the limit of large $\alpha$ the
random collisions occur effectively only between nearest neighbours
\begin{equation} \label{Wnn}
  \lim_{\alpha\rightarrow\infty} \mathcal{W}_\alpha(r) = \delta_{r,1}
  \ .
\end{equation}
Also, the long-range interacting harmonic chain with nearest-neighbour
random collisions was studied in \cite{tamaki2020}.  Furthermore, if
we also consider the limit of large $\delta$, then the model reduces
to a harmonic chain with nearest-neighbour random collisions studied in
\cite{lepri2009} and originally introduced in \cite{basile2006}. 
The dynamics with finite $\alpha$ has not been considered in the
past. As already remarked in Ref.~\cite{tamaki2020} for values $\delta <2$
correlations are ill-defined in the thermodynamics limit. We will show
below that we find similar pathological behaviors for finite $\alpha$ in the
same range of $\delta$. Therefore, we restrict our analysis to $\delta \ge 2$.

\section{Equations of motion}
\label{sec:dyn}

In this section we describe the dynamics of Hamiltonian (\ref{H})
perturbed by a long-range conservative noise. Taking advantage of
the periodicity of the system, the dynamics can be naturally cast in
the basis of Fourier normal modes (see \ref{sec:fourier}).  This
program was recently followed in Ref.~\cite{lepri2023} for 
a general harmonic network with stochastic collisions between
oscillators.  Here, we apply this approach to a network with long-range
couplings and we extend it to case in which a long-range conservative noise is present (see eq.~\ref{Wprob}) .


To  start  with,  we  note  that  the  long-range  interaction  between
oscillators can be written as
\begin{equation} \label{phi}
  \frac{1 }{N_\delta} \sum_{i=1}^N \sum_{r=1}^{N/2} \frac{(q_{i+r} -
    q_i )^2}{2 r^\delta} = \frac{1}{2}\sum_{i,j=1}^N q_i
  \Phi_{i j} q_j \ ,
\end{equation}
where the interaction matrix $\Phi$ is given by
\begin{equation} \label{phi-matrix}
  \Phi_{i j}=
  2\delta_{ij}-\frac{1}{N_{\delta}}\left(\min(|i-j|,N-|i-j|)\right)^{-\delta}
  \ .
\end{equation}
In the limit of large $\delta$, $\Phi$ reduces to the well known discrete
Laplacian describing the interaction term of the Harmonic chain.
Since $\Phi_{i j}$ is a circulant
matrix, it is diagonalised by Fourier normal modes \cite{davis1979},
namely $\Phi \chi^\nu = \omega_\nu^2 \chi^\nu$, with eigenvectors
$\chi^\nu_l$ as given in Eq.~\ref{chi}, and eigenvalues
$\omega_\nu = \omega(k_\nu)$ given explicitly by
\begin{equation} \label{omega}
 \omega_\nu = \frac{1}{\sqrt{N_\delta}} \left( \sum_{r=1}^{N/2}
   \frac{4\sin^2\left(\frac{k_\nu r}{2}\right)}{r^\delta}\right)^{1/2} \ ,
\end{equation}
being their corresponding frequencies.
For our purposes,  it is useful to write down the expression of 
spectrum valid in the limit $N\rightarrow\infty$.  
Using Euler's formula we can express \ref{omega} as
\begin{equation} \label{w-Li}
  \omega_{\nu}^{2}= \frac{1}{\zeta(\delta)}\left(2\zeta(\delta) -
    \Li_\delta\left(e^{i k_{\nu}}\right)  - \Li_\delta\left(e^{-i k_{\nu}}\right)
  \right) \ ,
\end{equation}
where $L_s(z) \equiv \sum_{r=1}^\infty (z^r/r^s)$ is the polylogarithm
function.

Therefore, in normal mode coordinates the Hamiltonian \eref{H}
acquires the simple form
\begin{equation} \label{F-H}
  H(\mathbf{Q},\mathbf{P}) = \frac{1}{2} \sum_{\nu} \left( \left|
      P_\nu \right|^2 + \omega_\nu^2 \left|
      Q_\nu \right|^2\right) \ .
\end{equation}
Using the approach of Ref.~\cite{lepri2023} one can 
compute exactly 
the change in normal mode momenta due to collisions between oscillators $m$ and $n$ as      
\begin{equation} \label{colrule}
  \mathbf{P}^\prime = \mathbf{P} - 2 \mathbf{V} \mathbf{V}^\top
  \mathbf{P} \ ,
\end{equation}
where  $\mathbf{P}^\prime$ denotes  the normal  mode momenta  after the
collision, and $\mathbf{V}=\mathbf{V}^{(n,m)}$ is a vector with components
\begin{equation} \label{V}
V_\nu^{(n,m)} = \frac{\chi_n^\nu - \chi_m^\nu}{\sqrt{2}} \ .
\end{equation}

In terms of coordinates $A_\nu$ defined in Eq.~\ref{A}, the collision
rule can be written as \cite{lepri2023}
\begin{equation} \label{DA}
  \mathbf{A}^\prime = \mathbf{A} - \mathbf{M}\left(\mathbf{A} +
    \mathbf{A}^*\right) \ ,
\end{equation}
where the matrix $\mathbf{M} = \Omega^{-1/2}\mathbf{V} \mathbf{V}^\top
\Omega^{1/2}$, and $\Omega$ the frequency matrix \eref{Omega}.

Together   with  the   deterministic  evolution   of  Eq,~\ref{Afree},
Eq.~\ref{DA}  determines   the  fundamental  evolution  step   of  our
system.  To write  it  in compact  form let  us  define the  auxiliary
$N$-dimensional vectors
\begin{equation} \label {vectors}
U_\nu = \sqrt{\frac{2}{\omega_\nu}} \ V_\nu \ , \quad W_\nu =
\sqrt{2\omega_\nu} \ V_\nu \ ,
\end{equation}
in terms of which \eref{DA} can be written as 
\begin{equation} \label {DA2}
\mathbf{A}^\prime = \mathbf{A} - \mathbf{U}\left(
  \mathbf{W}^\dagger\mathbf{A} + \mathbf{W}^\top\mathbf{A}^*\right) \ .
\end{equation}

Combining \eref{Afree} and \eref{DA2} the fundamental evolution map of
$\mathbf{A}$ from time $t$ immediately after a collision to time $t+\tau$
immediately after the subsequent collision is
\begin{equation} \label{Aevol}
  \mathbf{A}(t+\tau) = (1 - \mathbf{U} \mathbf{W}^\dagger) e^{i\Omega\tau} \mathbf{A} -
  \mathbf{U} \mathbf{W}^\top e^{-i\Omega\tau} \mathbf{A}^* \ .
\end{equation}

Note that  the randomness of  the stochastic collisions  is implicitly
carried on  in the vector $\mathbf{V}^{(n,m)}$,  where oscillators $n$
and $m$ are  randomly chosen. This is like choosing  a specific vector
$\mathbf{V}$ out  of a pool  of $N(N-1)/2$ different vectors,  each of
which correspond to the randomly chosen couple $n$, $m$ of oscillators
to exchange momentum.

The advantage  of using the  coordinates vector $\mathbf{A}$  is that,
once  the  collision between  two  oscillators  has been  picked,  the
evolution step  \eref{Aevol}, is obtained simply  by multiplication of
$N$-dimensional vectors, which is numerically quite efficient.

\section{Autocorrelation of the total energy current}
\label{sec:cor}

The autocorrelation of the energy current at equilibrium is of major 
importance for heat transport, since it enters in the Green-Kubo
formula for the conductivity.
As it is customary, the energy current is defined by means of the
continuity equation of the energy. 

\subsection{Energy currents}

For our system,  the current
has two contributions: one coming from the deterministic dynamics
generated by the Hamiltonian \eref{H}, and the other from the
stochastic noise generated by the random binary collisions.
For  the deterministic  contribution,  we  first
rewrite    Eq.~(\ref{H})     as    a     sum    of     local    energies
$H = \sum_{i=1}^N h_i $, where
\begin{equation} \label{hi}
h_i = \frac{p_{i}^{2}}{2} +
  \frac{1 }{2N_\delta} \sum_{r=1}^{N/2} \left(\frac{(q_{i+r} -
    q_i )^2}{2 r^\delta} + \frac{(q_{i-r} -
    q_i )^2}{2 r^\delta}\right) \ ,
\end{equation}
and identify  the local  current $j_i$  from a  discretised continuity
equation  
$\d h_i/\d t = -\left(j_i - j_{i-1}\right)$.  We obtain
\begin{equation}  \label{Jdet}
    j^{(det)}_{i}= -\frac{1}{N_{\delta}}\sum_{m=i+1}^{i+N/2}
    \sum_{r=i-m}^{N/2} \frac{(q_{m}-q_{m-r})(p_{m}+p_{m-r})}{2r^{\delta}} \ ,
\end{equation}
where we  use the label  $(det)$ to identify Eq.~(\ref{Jdet})  as purely
deterministic. Note that  the limit $\delta\rightarrow\infty$ recovers
the  expression  of   the  energy  current  for   the  harmonic  chain
\cite{lepri2003}.

The second  contribution to  the total  energy current  is due  to the
random  collisions.  They yield a  infinitesimal stochastic evolution given by
\begin{eqnarray} 
 \fl      dq_{i}&=p_{i}dt, \\
  \fl      dp_{i}&=\displaystyle
          \frac{1}{N_{\delta}}\sum_{r=1}^{N/2}\frac{q_{i+r}-2q_{i}+q_{i-r}}{r^{\delta}}
          \d t + \sum_{r=1}^{N/2}\left(\d n_{i,i+r}
        (p_{i+r}-p_{i})+\d n_{i,i-r}(p_{i-r}-p_{i})\right) \ , \label{stoEvol}
\end{eqnarray}
where $\d n_{i,i'}$ are random Poisson variables which can be either
$0$ or $1$ with average $\langle \d n_{i,i'}\rangle = \gamma
|i-i'|^{-\alpha}\d t$.  The first term of the \emph{r.h.s} of
\eref{stoEvol}  is reminiscent of the discrete fractional Laplacian.
Indeed, it reduces to the standard discrete Laplacian for $\alpha \to \infty$, matching 
diffusion in momentum space as described in Refs.~\cite{basile2006,tamaki2020}. 

Using Eq.~(\ref{stoEvol}) we derive the  change of the respective energy
density and, as before, using the continuity  equation, it turns out  that the
evolution of the stochastic contribution to the energy current has terms 
\begin{equation} \label{eq:djsto}
  \d j_i^{(sto)} = j_i^{(S)} \d t + \d \mathfrak{j}_i \ ,
\end{equation}
where 
\begin{equation} \label{eq:jsa}
    j^{(S)}_{i} = -\frac{1}{N_{\alpha}}\sum_{r=1}^{N/2} \left(
    \frac{p_{i+r}^{2}-p_{i}^{2}}{2}\right) \ ,
\end{equation}
is due to the energy exchange during the collisions, and
\begin{equation} \label{eq:jsb}
  \d\mathfrak{j}_{i}=-\frac{1}{N_{\alpha}} \sum_{r=1}^{N/2}
  \d\mathfrak{m}_{i,i+r}\left(
    \frac{p_{i+r}^{2}-p_{i}^{2}}{2}\right) \ ,
\end{equation}
is purely due to the noise. Here $\d\mathfrak{m}$ denotes the fluctuations
of the process $\d n$ around its average 
\begin{equation} \label{eq:martigale}
 \d\mathfrak{m}_{i,i+r} = \d n_{i,i+r}-\langle\d n_{i,i+r}\rangle \ .
\end{equation}
The limit  $\alpha\rightarrow\infty$ correctly recovers  the evolution
with nearest-neighbour collisions studied in \cite{basile2006,tamaki2020}.
Summing over sites, the total energy current reduces to
\begin{equation} \label{eq:Jtot}
  J_N = J_N^{(det)} + \d\mathfrak{J}_N \ , \ \mathrm{with} \quad
   J_N^{(det)} = \sum_{i=1}^N j_i^{(det)} \ , \ \
  \d\mathfrak{J}_{N} = \sum_{i=1}^N \d\mathfrak{j}_i \ .
\end{equation}
This  follows from  realising  that the  sum  $\sum_i j^{(S)}_{i}$  is
telescopic and thus, over the sum, is identically zero.

Therefore, the autocorrelation of the total current contains the
following contributions
\begin{equation} \label{CN0}
\fl \mathcal{C}_N(t) = \frac{1}{N}\left( \langle J_N^{(det)}(t) J_N^{(det)} \rangle + \langle
J_N^{(det)}(t) \d\mathfrak{J}_N
\rangle + \langle \d\mathfrak{J}_N(t) J_N^{(det)} \rangle + \langle
\d\mathfrak{J}_N(t) \d\mathfrak{J}_N \rangle \right) \ .
\end{equation}
The    cross   correlation    terms   involving    $J_N^{(det)}$   and
$\d\mathfrak{J}_N$ yield  a vanishing  contribution as the  current is
odd with respect to time reversal ~\cite{basile2006}.

Concerning the noise contribution $\d\mathfrak{J}_N(t)
\d\mathfrak{J}_N$, it was shown in Ref.~\cite{basile2006} that when
the noise applies only to  nearest-neighbour oscillators ($\alpha =
\infty$), this term is to leading order
\begin{equation}
  \langle \d\mathfrak{J}_N(t) \d\mathfrak{J}_N \rangle \approx
  \frac{\gamma}{N} \ ,
  \label{noisecont}
\end{equation}
and thus, negligible for large $N$. In the case 
long-range deterministic
interactions it also neglected \cite{tamaki2020}.
For the present case, we will comment
on the effect of such term later on in Section \ref{sec:heatcond}.
%



For the sake of clarity,  in what  follows we  will  denote  the
deterministic   contribution    to   the    autocorrelation   function
$\mathcal{C}_N(t)$  simply as  $C_N(t)$, and   the  noise contribution  as
$\mathfrak{C}_N(t) \equiv \langle \d\mathfrak{J}_N(t) \d\mathfrak{J}_N
\rangle$.

\subsection{Autocorrelation of the deterministic current}

We focus on the deterministic contribution to the autocorrelation
function of the total energy current $ J_N^{(det)}$, defined as
\begin{equation} \label{CNdet}
C_N(t)\equiv  \frac{1}{N} \langle J_N^{(det)}(t) J_N^{(det)} \rangle \ .
\end{equation}
Decomposing into normal modes, the total energy current can be written as
\cite{lepri2003}
\begin{equation} \label{J}
  J_N^{(det)} =\sum_{\nu} v_{\nu} E_{\nu} = \sum_{\nu}v_{\nu} \,  \delta E_{\nu}
\end{equation}
where 
\begin{equation} \label{E}
  E_\nu(t) = \omega_\nu \left| A_\nu(t) \right|^2; \quad   
  \delta E_\nu(t) = E_\nu(t) - \langle
      E_{\mu}\rangle_{\mathrm{eq}}
\end{equation}
are, respectively,  the energy of the $\nu$-th normal mode 
and its deviation from the equilibrium values $ \langle
      E_{\mu}\rangle_{\mathrm{eq}}$ and  $v_\nu=v(k_\nu)\equiv \partial_{k_\nu} \omega(k_\nu)$,  its group velocity as given explicitly from (\ref{omega})
\begin{equation} \label{v}
 v_\nu = \frac{1}{N_\delta\omega_\nu} \sum_{r=1}^{N/2}
   \frac{\sin\left(k_\nu r\right)}{r^{\delta-1}} \ .
\end{equation}


A remarkable property of this type of models is that,  
in  the kinetic  limit (corresponding  to time  scales  on which  each
oscillator  has  suffered  at  least one  collision  on  average ) 
the mode energy obeys a linear master equation, see (\ref{Eevol})
in \ref{sec:kin} \cite{lepri2023}.  
The transition 
rates can be computed explicitely from the eigenvectors $\chi_\nu$
upon averaging over the distribution of 
collision probabilities (some details are in \ref{sec:kin}). 
Therefore, relaxation to equilibrium is controlled by 
the characteristic rates $\mu_{\nu}$ 
that can be computed as 
eigenvalues of the master equation itself \cite{lepri2023,lepri2024large}.
Altogether,  if  the initial state  is at equilibrium at  temperature $T$,
$\left\langle    \delta E_{\mu}(0)   \delta E_{\nu}(0)\right\rangle    =   2\left(\kB
  T\right)^2  \delta_{\mu,\nu}$,   with  $\delta_{\mu,\nu}$   is  the
Kronecker delta function, and the autocorrelation function of the
total energy current becomes a
function of the group velocity and the relaxation rate of the energy
modes
\begin{equation} \label{CNt}
  C_{N}(t) = \frac{2\left(\kB T\right)^2}{N}
  \sum_{\nu}v_{\nu}^2 \ e^{-|\mu_{\nu}|t} \ .
\end{equation}

In the present case, the chain is translationally invariant and the 
the operator entering the master equation is almost diagonal up to $O(1/N)$ corrections, 
see (\ref{Rmatrix}).  Thus,  
one can approximate the $\mu$ by the diagonal elements as
\begin{equation} \label{mu}
  \mu_{\nu} \approx -\frac{4\gamma}{N_\alpha} \sum_{l=1}^N
  \frac{\sin^{2}\left(k_{\nu}l/2\right)}{l^\alpha} \ , \quad N_\alpha
  = \sum_{k=1}^{N/2} 1/l^\alpha \ ,
\end{equation}
where $\gamma$  is the rate at  which random collisions occur,  and is
defined in  the kinetic limit in  Eq.~(\ref{gamma}).
In the  kinetic limit,  we require that measurements  are taken  on time
scales   larger  than   the  application   of  the   fundamental  step
\eref{Aevol} $N$  times. On this  time-scale the effect of  the random
collisions                   becomes                     macroscopic
\cite{delfini2010,Iacobucci2010,lepri2020,lepri2023}.
The accuracy of the exponential  relaxation of  the energy modes  
and approximation (\ref{mu})  was
tested  numerically in  \cite{lepri2023}  
finding a very good agreement.

Let us consider now the thermodynamic limit $N\rightarrow\infty$. 
Upon replacing 
$k_\nu= 2\pi \nu/N$ with a continous variable $k_\nu\to k$, 
and $\mu_\nu\to \mu(k)$, the sum in Eq.~(\ref{CNt}) becomes an integral over the
momentum $k$
\begin{equation} \label{Ccont}
  C_{N}(t) = \frac{2(k_{B}T)^{2}}{2\pi}\int_{\frac{2\pi}{N}}^{2\pi} \d
  k \,v^{2}(k)
  e^{-|\mu(k)|t} \ .
\end{equation}
In the large time limit $t\rightarrow\infty$, the integral in
Eq.~(\ref{Ccont}) is asymptotically dominated by the low momenta
$k$. Therefore, we can extend the upper limit of the integral to
$\infty$, committing an exponentially small error
\begin{equation} \label{C-2}
  C_{N}(t) = \frac{(k_{B}T)^{2}}{\pi}\int_{\frac{2\pi}{N}}^{\infty} \d
  k\, v^{2}(k) e^{-|\mu (k)|t} \ .
\end{equation}
It is important to remark that we do not send to zero the lower extremum of the integral in Eq.~(\ref{C-2}) because it encodes the correction for large, but finite, $N$ to the correlation function. As we will see, these corrections are 
essential to match our analytical predictions with the numerical data.

\section{Analytical results}

We are now in the position to find the asymptotic approximations of
expression (\ref{C-2}). To this aim we need to assess the small
wavenumber behavior of the integrand.

\subsection{Small wavenumber limits}

To obtain the limit of low $k_\nu$, corresponding to the long time
limit, one can use  the following power-series  expansion of the
polylogarithm \cite{wood1992}
\begin{equation} 
\label{exp}
  \Li_{s}(e^{z}) = \Gamma(1-s)(-z)^{s-1} + \sum_{m=0}^{\infty}
  \frac{\zeta(s-m)}{m!}z^{m} \  ,
\end {equation}
valid for $|z| < 2\pi$ and $s \notin \mathbb{N} $ or:
\begin{equation}
    \label{expint}
 \Li_{s}(e^{z}) = \frac{z^{s-1}}{(s-1)!}[\sum_{l=1}^{s-1} \frac{1}{l}-\ln(-z)] + \sum_{m=0, m\neq s-1}^{\infty}
  \frac{\zeta(s-m)}{m!}z^{m}, 
\end{equation}
which is valid for $s\in\mathbb{N}$. In the following we will consider only non-integer values for $\alpha$ and $\delta$.
Keeping the leading-order terms in the expansion, this leads to
the small-wavenumber asymptotics for the spectrum
\begin{equation}
  \omega^2(k) = \left\{
    \begin{array}{ll}
      a_\delta |k|^{\delta-1}  \ , & 1 < \delta <3 \ , \\
      \\
      b_\delta k^2 \ , &  \delta > 3\ ,
    \end{array}
  \right.
\end {equation}
and for the group velocity 
\begin{equation}  \label{v-sol}
  v(k) = \left\{
    \begin{array}{ll}
      \left(\frac{1-\delta}{2}\right) a_\delta^{1/2} |k|^{(\delta-3)/2}  \ , & 1 < \delta <3 \ , \\
      \\
      b_\delta^{1/2} \mathrm{sign}(k) \ , &  \delta > 3\ ,
    \end{array}
  \right.
\end {equation}
with
\begin{equation} \label{constants}
  	a_{s} = -\frac{2}{\zeta(s)} \Gamma(1-s)\sin\left(\frac{\pi
            s}{2}\right) \ , \qquad b_{s} = \frac{\zeta(s-2)}{\zeta(s)}.
\end{equation}

Finally,  a similar calculation for  the decay rates $\mu(k)$ (\ref{mu}) yields  
\begin{equation} 
\label{mu-sol}
  \mu(k)= \left\{
    \begin{array}{ll}
      \gamma a_\alpha |k|^{\alpha-1}  \ , & 1 < \alpha <3 \ , \\
      \\
      \gamma b_\alpha k^2 \ , &  \alpha > 3\ .
    \end{array}
  \right.
\end {equation}

We see that Eqs.~(\ref{v-sol}) and (\ref{mu-sol}) hint at four different
regimes for the autocorrelation function Eq.~(\ref{C-2}) classified by
whether the interaction between oscillators is effectively short or
long ranged, and whether the random collisions regarding as the
conservative noise is short or long ranged.

When interactions  and collisions  are both short  range, specifically
affecting only  nearest neighbours, the  system results in  a harmonic
chain with conservative noise introduced in Ref.~\cite{basile2006} and
further studied in \cite{lepri2009,lepri2010,tamaki2020}. The chain of
long-range interacting oscillators with conservative noise was studied
in   Ref.~\cite{tamaki2020},  and   analytical  expressions   for  the
asymptotic behaviour of $C_N(t)$. The last two regimes, not previously
studied,  correspond to  add  long-range random  collisions, namely  a
long-range   interacting   chain   of  oscillators   with   long-range
conservative noise, and a short-range harmonic chain with a long-range
conservative noise.

\subsection{Decay in the thermodynamic limit}

Using the low-$k$ limit expressions for the group velocity and
relaxation rate of the previous section, the total energy correlation
(\ref{C-2}) can be synthetically written as
\begin{equation} \label{C-3}
  C_N(t) = \frac{(\kB
    T)^2}{\pi}\mathcal{Z}(\delta)\int_{\frac{2\pi}{N}}^\infty k^p
  e^{-\gamma^\prime(\alpha) t \ k^q} \d k \ ,
\end{equation}
where
\begin{equation} \label{pq}
  p \equiv p(\delta) = \left\{
    \begin{array}{ll}
      \delta-3 \ , & \!\! 1 < \delta < 3 \\ 
      \\
      0 \ , & \!\! \delta > 3
    \end{array}
  \right. \!\! \!\!  ,
  \quad
    q \equiv q(\alpha) = \left\{
    \begin{array}{ll}
      \alpha-1 \ , & \!\! 1 < \alpha < 3 \\ 
      \\
      2 \ , & \!\! \alpha > 3
    \end{array}
  \right. \!\! \!\!  ,
\end{equation}
and
\begin{equation} \label{Z}
  \mathcal{Z}(\delta) = \left\{
    \begin{array}{ll}
      a_\delta \left(\frac{\delta-1}{2}\right)^2 \ , & \!\! 1 < \delta < 3 \\ 
      \\
      b_\delta \ , & \!\! \delta > 3
    \end{array}
  \right. \!\! \!\!  ,
  \quad 
    \gamma^\prime(\alpha) = \left\{
    \begin{array}{ll}
      a_\alpha\gamma \ , & \!\! 1 < \alpha < 3 \\ 
      \\
      b_\alpha\gamma \ , & \!\! \alpha > 3
    \end{array}
  \right.  \!\! \!\! .
\end{equation}
As it may intuitively expected, the nominal collision rate 
$\gamma$ is rescaled by the parameter $\alpha$ determining
the range of the collisions.  

It is easy to see that the correlation function Eq.~(\ref{C-2}) diverges for $N\rightarrow\infty$ if $1<\delta<2$. Indeed, the divergence is due the low values of $k$: in this region the exponential is immaterial and so the behaviour of the correlation function is the same as the one reported in ~\cite{tamaki2020}, which is divergent in $N$ for $1\le \delta \le 2$. 
Since in this range of parameters the thermodynamic limit is not well defined and thus we will not consider it and restrict ourselves to the range $1<\delta<3$.
For completeness,  we explicitly compute Eq.~(\ref{C-2}) for the case $\delta=2$ 
in \ref{sec:delta2}. 

Equation (\ref{C-3}) can be integrated to give
\begin{equation} \label{Ctheory}
  C_N(t) = \frac{(\kB
    T)^2}{\pi}\frac{\mathcal{Z}(\delta)}{q(\alpha)}
  \left(\gamma^\prime t\right)^{-(1+p(\delta))/q(\alpha)}
  \Gamma\left[\frac{1+p(\delta)}{q(\alpha)},
      \left(\frac{2\pi}{N}\right)^{q(\alpha)} \gamma^\prime t \right]
    \ ,
\end{equation}
where $\Gamma[a,x]$ is the incomplete Gamma function defined as
\begin{equation} \label{Gamma}
  \Gamma[c,z] = \int_z^\infty t^{c-1} e^{-t} \d t \ .
\end{equation}

Equation~\ref{Ctheory} constitutes our main 
analytical result. It reveals that in the
thermodynamic limit $N\rightarrow\infty$, the autocorrelation function
of the total energy current $C_\infty(t)\equiv\lim_{N\to \infty}C_N(t)$
 decays as a power-law in time, specifically
\begin{equation} \label{C-thermlimit}
 C_\infty(t) \sim t^{-\beta},\qquad \beta\equiv{\frac{1+p(\delta)}{q(\alpha)}}
\end{equation}
If $1+p\le q$ the Green-Kubo integral diverges and
we have anomalous heat transport. We will discuss this in detail later on.

\subsection{Finite-$N$ effects}

In contrast, for any finite $N$, the autocorrelation function and does not decay as a power
law due to the residual dependence on $N$ of the incomplete $\Gamma$ function.
The characteristic scaling for the autocorrelation decay is
$\gamma^\prime t/N^{q(\alpha)}$. For
$\gamma^\prime t \ll N^{q(\alpha)}$ $C_N(t)$ decays as in
Eq.~(\ref{C-thermlimit}), while for $\gamma^\prime t \gg N^{q(\alpha)}$
the decay is dominated by the $\Gamma$ function which is asymptotically
exponential.

Furthermore, Eq.~(\ref{Ctheory}) admits a power expansion in $\gamma^\prime
t/N^{q(\alpha)}$, yielding
\begin{equation} \label{Cseries}
  \fl C_N(t) = \frac{(\kB
    T)^2}{\pi}\mathcal{Z}(\delta)
  \left[\frac{1}{q}\Gamma\left(\frac{1+p}{q}\right)\left(\gamma^\prime t\right)^{-\frac{1+p}{q}}  +
    \left(\frac{2\pi}{N}\right)^{1+p} \sum_{m=0}^\infty c_m(\delta,\alpha)
    \left(\frac{\gamma^\prime t}{N^{q}}\right)^m \right] \ ,
\end{equation}
with coefficients
\begin{equation} \label{cms}
  c_m(\delta,\alpha)
  =\frac{(-1)^{m+1}(2\pi)^{mq(\alpha)}}{m!(1+p(\delta)+mq(\alpha))}  \ .
\end{equation}
For  the sake  of brevity,  in Eq.~(\ref{Cseries})  we have  omitted the
dependence of $p$ and $q$.

Equation (\ref{Cseries}) is interesting  in that the thermodynamic limit
behaviour of the autocorrelation  function splits from the corrections
due to a  finite system size. This  is accounted by the  first term on
the  right-hand  side of  Eq.~(\ref{Cseries})  which  solely depends  on
time.  All size  dependence is  contained in  the second  term on  the
right-hand side of Eq.~(\ref{Cseries}).

More interestingly, the leading order of the series expansion, given by
\begin{equation} \label{leading}
  - \frac{(\kB
    T)^2}{\pi}\frac{\mathcal{Z}(\delta)}{1+p(\delta)}
  \left(\frac{2\pi}{N}\right)^{1+p(\delta)} \ ,
\end{equation}
is time  independent and,  perhaps strikingly, it  does not  depend on
$\alpha$. As a consequence, for any chain of finite size,
the autocorrelation function 
does not decay to zero for $t\to \infty$. 

\subsection{Transport regimes}

Based on the analytical results we can now distingush four
distinct transport regimes, depending on the role of
deterministic or stochastic energy transfers.
Taking into account the explicit expressions of $p,q$,  the asymptotic
scaling of the autocorrelation function in
the thermodynamic limit, expression  (\ref{C-thermlimit}),  can be made explicit as
\begin{equation} \label{Csummary}
  C_\infty(t) \sim \left\{
  \begin{array}{ll}
    t^{-1/2} \ , & I: \delta > 3\ , \alpha > 3 \\ 
    t^{-(\delta/2-1)} \ , & II: 2 < \delta < 3 \ ,
    \alpha > 3\\
    t^{-(\delta-2)/(\alpha-1)} \ , & III: 2 < \delta < 3 \ ,
    1 < \alpha < 3\\
    t^{-1/(\alpha-1)} \ , & IV: \delta > 3 \ ,
    1 < \alpha < 3 \ .
  \end{array} \right.
\end{equation}
We thus have  four regimes labeled by $I-IV$,  where 
the decay  of the autocorrelation depends  on the
effective range of  both the deterministic interaction  and the random
collisions, through the exponents $\delta$  and $\alpha$.  In the  case of long-range
interactions and  long-range collisions, the two  processes compete in
determining the decay of the correlation.  As we will discuss shortly, 
cases $I,II$ encompass the the known case of nearest-neighbour collisions
\cite{basile2006,tamaki2020} but we disclose   two
more regimes ($III,IV$) that have not been considered previously. 

\begin{itemize}

\item \textit{Regime I: Short-short range,  $\delta > 3\ , \alpha > 3$}. Here, 
the decay is the same as the 
for nearest-neighborur   harmonic chain with nearest-neighborur collisions
($\delta=\alpha=\infty$).  Indeed, Eq.~(\ref{Ctheory}) reproduces the
asymptotic results in Ref.~\cite{basile2006}, namely that
$\lim_{N\rightarrow\infty}\lim_{t\rightarrow\infty} C_N(t) \sim
t^{-1/2}$.  This means that interactions that falls sufficiently 
fast belong to the same universality class of anomalous transport.

\item \textit{Regime II: Short-long range,  $2 <\delta < 3, \alpha > 3$.}
Eq.~(\ref{Ctheory}) yields
\begin{equation} \label{CLS}
  C_N(t) = \frac{(\kB
    T)^2}{2\pi} \frac{a_\delta(\delta-1)^2}{4}
  \left(\gamma t\right)^{-\frac{\delta-2}{2}}
  \Gamma\left[\frac{\delta-2}{2},
      4\pi^2 \frac{\gamma t}{N^2} \right]
    \ .
\end{equation}
The decay of $C_\infty$, Eq.~(\ref{Ctheory}) agrees with the analytical
results obtained in Ref.~\cite{tamaki2020} for a harmonic long-range
interacting chain with nearest-neighbor collisions ($\alpha=\infty$)
in the regime $2 <\delta < 3$.
We stress that, In both cases $I,II$, our expression (\ref{Ctheory}) 
generalizes and improves the previous
results in that it determines the autocorrelation function for any
finite $N$ and $\alpha$, including the prefactors.

\item \textit{Regime III: Long-long range,  $2<\delta<3$,
$1<\alpha<3$}
%
In this regime,, the autocorrelation function becomes
(from Eq.~(\ref{Ctheory}))
\begin{equation} \label{CLL}
  C_N(t) = \frac{(\kB
    T)^2}{4\pi} \frac{a_\delta(\delta-1)^2}{\alpha-1}
 \left(a_\alpha \gamma \ t\right)^{-\frac{\delta-2}{\alpha-1}}
  \Gamma\left[\frac{\delta-2}{\alpha-1},
      4\pi^{\alpha-1} \frac{a_\alpha \gamma \ t}{N^{\alpha-1}} \right]
    \ .
\end{equation}
This is a novel regime, where the decay is determined by the interplay between 
both deterministic interactions and the random collisions. 
In fact the thermodynamic limit  the decay of the  autocorrelation function 
$C_\infty(t) \sim \left(\gamma \ t\right)^{-(\delta-2)(\alpha-1)}$
depends on both exponents.  As
discussed above, note also that for finite size chains, the
leading-order correction to this decay is
\begin{equation}
  - \frac{(\kB
    T)^2}{\pi}\frac{\mathcal{Z}(\delta)}{\delta-2}\left(\frac{2\pi}{N}\right)^{\delta-2}
  \ .
\end{equation}
In contrast with the decay of the autocorrelation in the thermodynamic
limit, the leading-order of the finite-size correction does only
depend on the effective range of the deterministic interaction. Note
as well that for $\delta\gtrsim 2$ the finite-size corrections decay
extremely slowly with $N$.

\item \textit{Regime IV: Long-short range,  $\delta>3$,
$1<\alpha<3$.} The last regime is the case 
where a long-range conservative noise dominates the decay,  and 
 Eq.~(\ref{Ctheory}) yields
\begin{equation} \label{CSL}
  C_N(t) = \frac{(\kB T)^2}{\pi} \frac{b_\delta}{\alpha-1}
 \left(b_\alpha \gamma \ t\right)^{-\frac{1}{\alpha-1}}
  \Gamma\left[\frac{1}{\alpha-1},
      4\pi^{\alpha-1} \frac{b_\alpha \gamma \ t}{N^{\alpha-1}} \right]
    \ .
\end{equation}
Since the random collisions can occur at distances as long as the size
of the chain, it is not surprising that this regime is dominated by them
through its effective range $\alpha$.  Notably, the only contribution
of  the
deterministic dynamics is in the prefactor of
Eq.~(\ref{CSL}). 
\end{itemize}
A remarkable fact is that 
there are two subdomains of regions $III$ and $IV$, namely
\begin{equation}
\alpha < \delta-1, \quad 2<\delta<3 \quad (III) \qquad 
1<\alpha<2, \quad\delta>3  \quad (IV)
\label{normal}
\end{equation}
where $C$ decays faster than $1/t$. This suggest a normal diffusive transport
at least as far as the deterministic contribution to the current is 
considered. 

As a final note, it is worthwhile to remark that the comparison
between our analytical expressions with the numerically computed
autocorrelation is hindered by additional finite size corrections,
since Eq.~(\ref{Ctheory}) was derived in the limit of large system
sizes and large times. A careful discussion of this will be presented
in the next section.

%

\begin{figure} [!t]
\centerline{\includegraphics[width=0.65\textwidth]{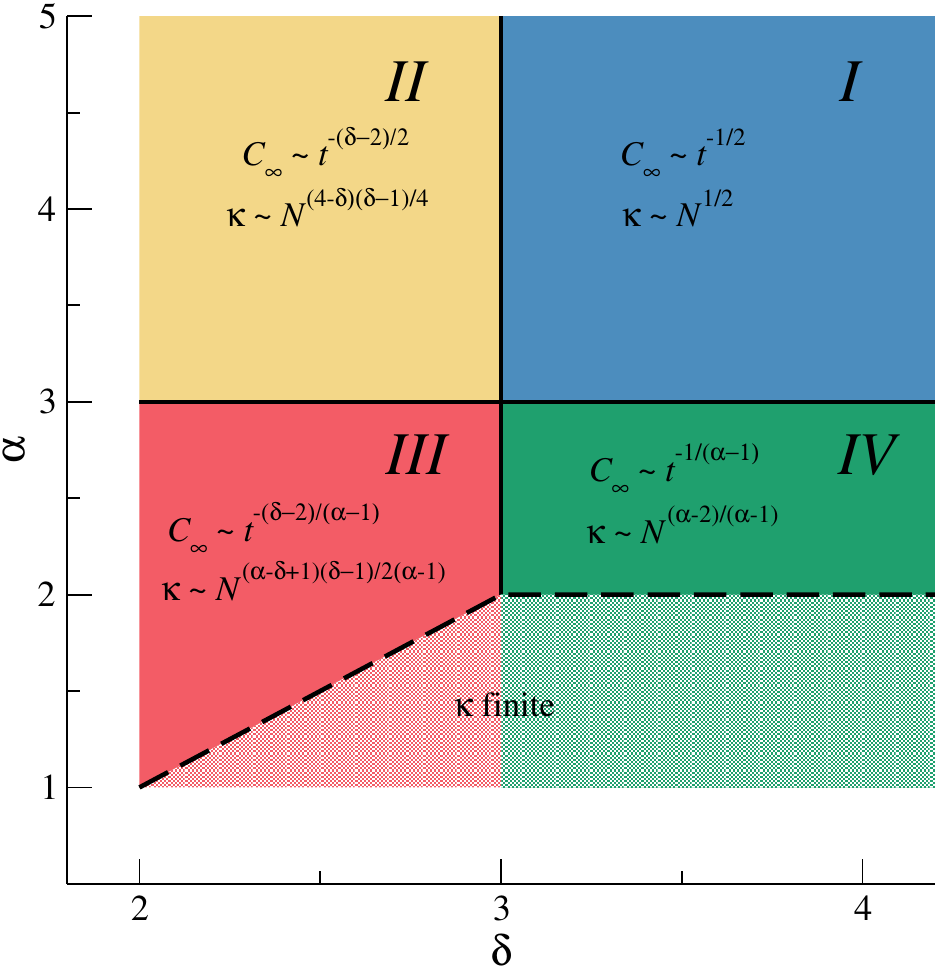}}
\caption{(Colour online) Phase diagram of the different regimes of
  the model system of section~\ref{sec:model}. Four different regimes
  are identified by colours and delimited by solid
  lines depending on the asymptotic decay of $C_\infty$. 
  The scaling of the thermal conductivity $\kappa$ with the
  system size $N$ is indicated by the respective labels. The region 
  below the dashed
  curve corresponds to the values of $\delta$ and $\alpha$ for which
  $\kappa$ does not scale with $N$, and thermal transport is
  diffusive (see the text for discussion).}
\label{fig:phases}
\end{figure}
\subsection{Heat conductivity} \label{sec:heatcond}

For  finite chains,  the
leading-order of the finite-size corrections  depend on $N$ but not on
$t$. Strictly  speaking, this means that  the autocorrelation function
exhibits a power-law decay only in thermodynamic limit $N=\infty$. For
large  but finite  systems, the  autocorrelation of  the total  energy
current  is not  a power  law.   Interestingly, the  amplitude of  the
leading-order correction is sensitive to  the value of $\delta$ but it
does not depend on $\alpha$.

The space-time scaling variable used to expand the autocorrelation in a
power-series is determined by the relaxation rate $\mu(k)$, which in
turns is set by the effective range of the random collisions $\alpha$.

The  four regimes  that our  dynamics  cover, are  characterised by  a
distinctive thermal transport which can be determined through the
Green-Kubo formula for the thermal conductivity given by
\begin{equation} \label{defG-K}
  \kappa = \frac{1}{\kB T^2} \lim_{t\rightarrow\infty}\lim_{N\rightarrow\infty}\int_0^t 
  C_{N}(t) \d t \ .
\end{equation}
Here, there is the crucial assumption that the decay of the 
full autocorrelation $\mathcal{C}$,  Eq.(\ref{CN0}) is dominated by $C_N$,
i.e. that the stochastic part is negligible for large $N$, as in the 
short-range case,  see again Eq. (\ref{noisecont}).
Note  also that  the  average in  Eq.~(\ref{CNdet})  is a  micro-canonical
average,  but  since  we  are considering  a  local  system,  ensemble
equivalence does hold and we can replace it with a canonical average.

An argument to estimate the finite-size thermal conductivity 
$\kappa(N)$ from  Eq.~(\ref{defG-K}) is to cutoff the integral as
\begin{equation} \label{G-K}
  \kappa(N) = \frac{1}{\kB T^2} \int_0^{t^{*}(N)} 
  C_{\infty}(t) \d t \ ,
\end{equation}
where  $t^{*}=N/v^{*}$  and 
$v^{*}$ is the  typical velocity of propagation  of energy excitations
\cite{lepri2003}. Basically, we  do integrate up to the  time in which
the  energy   has  effectively  propagated  through   the  chain.   In
particular, the introduction of this cut-off is necessary, as 
when transport  is   anomalous,  the  thermal  conductivity
diverges with  $N$ due  to the  fact that  the autocorrelation  of the
energy current has long-living power-law tails. 

We estimate $v_*$ as the maximal group velocity as given 
by (\ref{v-sol}) evaluated for the smallest possible wavenumber 
$v_*=v(k=2\pi/N)$. Then $v_*$ depends only on $\delta$ and is finite
for $\delta>3$, while it diverges as $N^{(3-\delta)/2}$ for  $1<\delta<3$
yielding
\begin{equation}
\label{tstar}
  t^{*}(N) \sim \left\{
    \begin{array}{ll} 
     N^{(\delta-1)/2}  \ , & 1 < \delta <3 \ , \\
      N \ , &  \delta > 3\ .
    \end{array}
  \right.
\end {equation}
Substituting this expression along with Eq.~(\ref{Csummary}) into (\ref{G-K}), and taking into account 
(\ref{normal}),  we obtain the predictions:
\begin{equation} \label{kappa}
  \kappa(N) \sim \left\{
  \begin{array}{ll}
    N^{1/2} \ , & I: \delta > 3\ , \alpha > 3 \\ 
    \textcolor{black}{N^{(4-\delta) (\delta-1)/4}} \ , & II: 2 < \delta < 3 \ ,
    \alpha > 3\\
    N^{(\alpha-\delta+1)(\delta-1)/2(\alpha-1)} \ , & III: 2 < \delta < 3 \ ,
    \delta -1 < \alpha < 3\\
    finite \ , & III: 2 < \delta < 3 \ ,
    1 < \alpha < \delta-1\\
    finite \ , & IV: \delta > 3 \ ,
    1 < \alpha < 2 \\
    N^{(\alpha-2)/(\alpha-1)} \ , & IV: \delta > 3 \ ,
    2 < \alpha < 3 \ .
  \end{array} \right.
\end{equation}
 
Therefore,  in the region below
the dotted line in Fig. \ref{fig:phases}) where the Green-Kubo integrand is convergent
suggesting normal diffusive transport. This is rather unexpected as
in such region a long-range exchange occurs which, intuitively, should
enhance transport. In region $IV$, where sound speed is finite,
an heuristic physical argument in support of this
can be traced back to the scaling of $\omega(k)$ and $\mu(k)$:
indeed for $\alpha>2$ (resp. for  $\alpha<2$) we have
$\omega(k)\gg\mu(k)$ (resp. $\omega(k)\ll\mu(k)$) for small $k$.
So, the waves are overdamped in the second case. This is consistent with
a diffusive behavior, similar to what seen for instance in the
coupled rotors models \cite{lepri2003}.
It should be however remarked that our analysis regards
only the deterministic part of the current.
A further analysis would be needed to test this prediction.

All our observations, including  our central result Eq.~(\ref{Ctheory}),
were obtained for the closed  system. We conjecture that the asymptotic
behaviour of the correlation functions and of the heat conductivity in
the different regimes have the  same asymptotic behaviour than an open
chain of  oscillators maintained  in a  nonequilibrium steady  state by
external thermostats.

\section{Numerical results}
\label{sec:num}

In    section~\ref{sec:dyn}    we    extended   the    method    in
Ref.~\cite{lepri2023} to  chains perturbed by  long-range conservative
noise,  resulting  in   the  stochastic  map  for   the  normal  modes
$\mathbf{A}$-coordinates of Eq.~(\ref{Aevol}),  that solves the hybrid
evolution  of  deterministic  harmonic  interactions  perturbed  by  a
stochastic  noise.  The method  is  computationally  convenient as  it
basically  amounts to  a sequence  of external  products among  vectors,
without  any  time-discretization  as  in  the  case  of  ordinary  or
stochastic differential equations.

At  time $t=0$,  we initialise  the chain at equilibrium state with
temperature $T$, as explained in  \ref{sec:equil}.
In the following  we set the temperature  to $\kB T =  1$, meaning that
the energy per normal mode is $1$.

The   system  evolves   through  the   repeated  application   of  the
Eq.~(\ref{Aevol}),  where  the  couple   of  oscillators  colliding  are
randomly  chosen  according  to  Eq.~(\ref{Wprob}) every  time.  In  the
kinetic    limit,    the   mean    collision    time    is   set    to
$\tau= (\gamma N)^{-1}$. To ensure  the validity of the kinetic limit,
in all simulations we have chosen to set the macroscopic time scale to
$4 N \tau$, meaning that in one time step, each oscillator suffers 8
collisions on average.

We have numerically computed the autocorrelation of the total energy
current defined in Eq.~(\ref{CNdet}) for chains of different sizes $N$
and several choices of the parameters $\delta$ and $\alpha$.
The purpose of this section is to confront our main result
Eq.~(\ref{Ctheory}) with these numerical results.

To start,  let  us first  discuss  the purely  short-range  situation by  setting
$\delta=\infty$ and  $\alpha=\infty$. This correspond to  the dynamics
of a harmonic chain of oscillators with nearest-neighbour interactions
collisions (regime $I$ in Fig.~\ref{fig:phases}) \cite{basile2006, lepri2009}. 
Using
Eqs.~(\ref{pq})       and      (\ref{Z}),       and      noting       that
$\lim_{\delta\rightarrow\infty}Z(\delta)        =       1$,        and
$\lim_{\alpha\rightarrow\infty}\gamma^\prime(\alpha)     =    \gamma$,
Eq.~(\ref{Ctheory}) becomes in the thermodynamic limit
\begin{equation}     \label{olla}
  C_\infty(t)     =     \frac{(\kB T)^2}{2\sqrt{\pi}} \left(\gamma t\right)^{-1/2} \ .
\end{equation}

\begin{figure} [!t]
  \centerline{\includegraphics[width=0.7\textwidth]{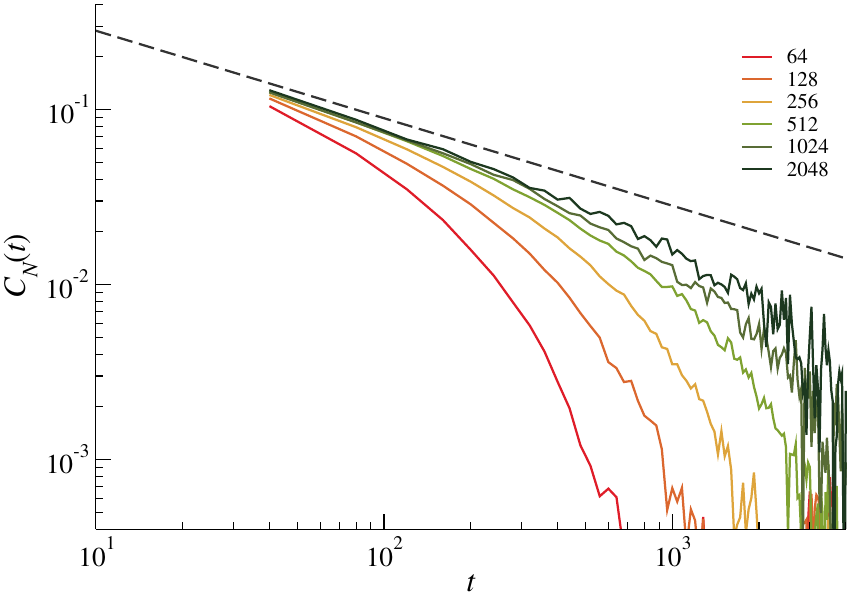}}
  \caption{Autocorrelation function $C_N(t)$ for the harmonic chain
    with conservative noise ($\delta=\infty$, $\alpha=\infty$) with
    $\gamma=0.1$, $n_{\mathrm{col}}=4N$ and size $N$ as indicated by
    the legends. The dashed curve corresponds to the theoretical
    expectation of Eq.~(\ref{olla}). }
\label{fig:Olla-vs-N}
\end{figure}

In Fig.~\ref{fig:Olla-vs-N}, we show the numerically computed
autocorrelation $C_N(t)$ for chains of different sizes. The dashed
curve corresponds Eq.~(\ref{olla}).  As evident from
Fig.~\ref{fig:Olla-vs-N}, the autocorrelation does converge to
the thermodynamic limit as $N\rightarrow\infty$.

However, the convergence is very slow  and not uniform. At short times
$t\approx10^2$,    the    convergence    is    well    described    by
$|C_N(t)  -   C_{th}|  \sim   1/\sqrt{N}$,  while  at   larger  times,
$t\approx 10^3$, the convergence is slower, approximately $\sim 1/N^3$.

It is worthwhile recalling that in this case, the noise contribution to the
autocorrelation function decays as $1/N$, see Eq.~(\ref{noisecont})
\cite{basile2006}, faster than the numerically observed rates of convergence.

Now, a  physically relevant  question is how  to determine  the asymptotic
decay in time of  the autocorrelation function. Equation~(\ref{Ctheory})
shows that the autocorrelation function  decays as a power-law of time
$C_N(t) \sim t^{-\beta}$, only if the chain is infinitely long. In the
thermodynamic  limit, the  power  exponent $\beta$  is  a function  of
$\delta$  and  $\alpha$  as  in Eq.~(\ref{Csummary}).  This  raises  the
question how estimate $\beta$ from numerical results of finite chains,
since a direct fitting of the data would yield inaccurate results.

\begin{figure} [!t]
  \centerline{\includegraphics[width=0.75\textwidth]{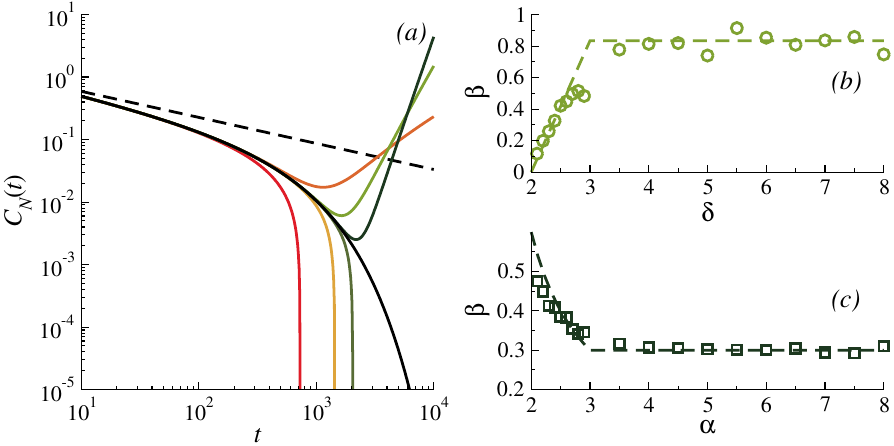}}
  \caption{Panel (a): Theoretical autocorrelation function $C_N(t)$
    for $\delta=2.5$, $\alpha=2.2$), $\gamma=0.1$,
    $n_{\mathrm{col}}=4N$ and $N=511$ (solid black curve). Coloured
    curves are successive approximations to order $m$, from $m=0$ (red)
    to $m= 5$ (dark blue).  The dashed curve corresponds to the
    thermodynamic limit $C_\infty(t))$. In panels (b-c) we show the
    decay exponent $\beta$ of $C_\infty(t))$ estimated from a fit to
    power law of the scaled autocorrelation $\tilde{C}_N(t)$ up to
    order $m=2$ for a chain of $N=511$ oscillators (symbols), compared
    to the theoretical expectation (dashed curves).  $\beta$ is shown
    as a function of $\delta$ for fixed $\alpha=2.2$ in panel (b) and
    as a function of $\alpha$ for fixed $\delta=2.6$ in panel (c).}
\label{fig:beta}
\end{figure}

The  crucial  observation is  that  in  the  series expansion  of  the
autocorrelation,  on the  right  hand side  of Eq.~(\ref{Cseries}),  the
dependence  on the  size of  the chain  appears only  in the  infinite
sum. Therefore, this  term can be considered a correction  term to the
asymptotic  decay in  time, accounting  for a  finite-size $N$  of the
chain.

In Fig.~\ref{fig:beta}($a$), we show an example of how, for a fixed
size $N=511$, the series expansion successively approaches the
autocorrelation $C_N(t)$ (black solid curve).  The dashed curve
correspond to the pure power-law decay of the first term of the right
hand side of Eq.~(\ref{Cseries}), namely the thermodynamic limit
$C_\infty(t)$.  The red curve corresponds to the
expansion~\eref{Cseries} up to leading order $m=0$.  As higher orders
are considered, Eq.~(\ref{Cseries}) describes well $C_N(t)$ to larger
and larger times (solid curves from red to dark blue). For instance,
we see that for the specific case shown in Fig.~\ref{fig:beta}
($\delta=2.6$, $\alpha=2.2$ and $N=511$), the series expansion up to
order $m=2$ (yellow curve), describes well $C_N(t)$ for times
$\lesssim 10^3$, which is the domain over which our numerical results
have a decent numerical convergence.

Noting the above we proceed as follows: From the numerically computed
autocorrelation $C_N(t)$ we subtract the value of the infinite sum of
Eq.~(\ref{Cseries}) up to second order $m=2$, specifically
\begin{equation} \label{Cscaled}
  \tilde{C}_N(t) = C_N(t) - 
  \left(\frac{2\pi}{N}\right)^{1+p} \sum_{m=0}^2
  \frac{(-1)^{m+1}(2\pi)^{mq(\alpha)}}{m!(1+p(\delta)+mq(\alpha))}
    \left(\frac{\gamma^\prime t}{N^{q}}\right)^m \ .
\end{equation}
Viewing the finite  size of the chain as a  correction, we obtain that
for all  values of $\delta$  and $\alpha$, the  scaling \eref{Cseries}
takes  the  numerically  computed  autocorrelation closer  to  a  pure
power-law of time.   Then we estimate the exponent $\beta$  from a fit
to  power-law  of  the   obtained  $\tilde{C}_N(t)$.  To  improve  the
estimation, we limit the power-law to the data at short times.

The results  are shown  in Fig.~\ref{fig:beta}($b$)  as a  function of
$\delta$  and  Fig.~\ref{fig:beta}($c$)   as  a  function  of $\alpha$.
Dashed  curves correspond  to  Eq.~(\ref{Csummary}).   The agreement  is
overall  good.    The  short-range   regimes  of  $\delta>3$   and  of
$\alpha>3$, are  in good agreement. Around  $\delta=3$, the estimation
of  $\beta$  becomes unstable  due  to  the divergences  of
\emph{e.g.},   $\mathcal{Z}(\delta)$.   The   agreement   is
reasonably good when the interactions  are long-range ($\delta <3$) or
when  the noise  is long-range  ($\alpha$). We argue that the
discrepancies observed are due to  the  fact that  Eq.~(\ref{Ctheory})
does not account for  the
noise term contribution to the autocorrelation. However, we cannot
discard the possibility of additional finite-size corrections due to
slow numerical convergence.

\begin{figure} [!t]
  \centerline{\includegraphics[width=0.8\textwidth]{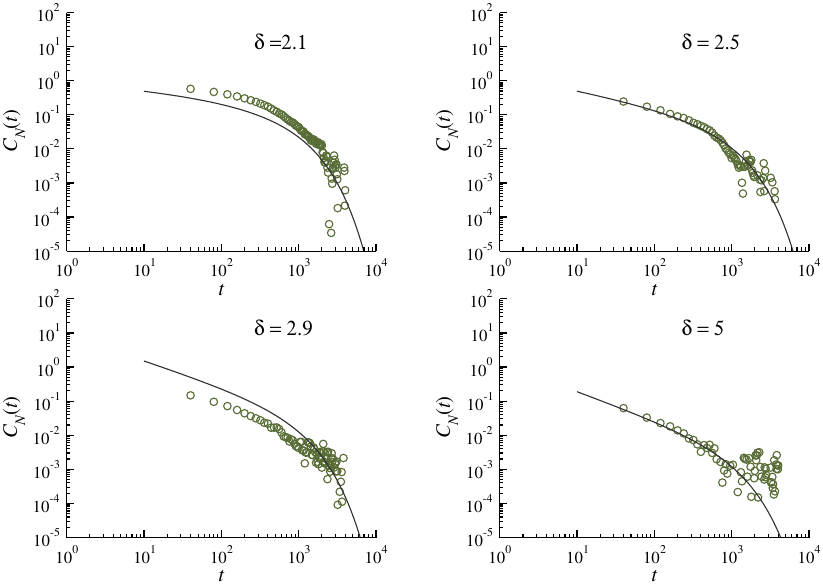}}
  \caption{Autocorrelation function of the total energy current
    $C_N(t)$ for $N=511$, $\alpha=2.2$, $\gamma=0.1$,
    $n_{\mathrm{col}}=4N$ and different values of $\delta$ (symbols).
    The curves corresponds to the theoretical expectations of
    Eq.~(\ref{Ctheory}). }
\label{fig:C-vs-delta}
\end{figure}

In Fig.~\ref{fig:C-vs-delta} we show the numerically computed
autocorrelation function $C_N(t)$ for a chain of $N=511$ oscillators,
$\alpha=2.2$ corresponding to random collisions that are effectively
long-range (99\% of the collisions occur between oscillators separated
a distance shorter than 27), and four different values of $\delta$.

The first  three figures,  with $2<\delta <  3$, correspond  to regime
$III$ long-range interacting chain  with long-range conservative noise
(red  region  in Fig.~\ref{fig:phases}).  There  is  a clear  mismatch
between the numerical results (circles) and Eq.~(\ref{Ctheory}) (solid
curves),  that depends  on the  value of  $\delta$. For  instance, for
values  of  $\delta\approx2.5$,  \eref{Ctheory}  seems  to  be  fairly
accurate.   However,  for $\delta  <  2.5$  our theory  underestimates
$C_N(t)$, while for $\delta > 2.5$, it overestimates it.

A closer inspection  of the numerical results and how  they compare to
Eq.~(\ref{Ctheory}),  shows that  the mismatch  is well  accounted by  a
$\delta$-dependent  global scaling  factor. For  instance, multiplying
the  autocorrelation  for  $\delta=2.1$  by  a  factor  $\approx0.45$,
minimises the quadratic distance between the data and the theory. Best
agreement  for $\delta=2.9$  is obtained  if we  multiply the  data by
$\approx2.58$.

Assuming that  these corrections can  be attributed to  neglecting the
contribution  of the  noise term  $\mathfrak{C}_N(t)$, the  results in
Fig.~\ref{fig:C-vs-delta}  suggest  that  while   in  the  regime  of
long-range   interactions  and   long-range  noise   $2<\delta<3$  and
$1<\alpha<3$, $\mathfrak{C}_N(t)$ is not  negligible, in the regime of
short-range interactions, $\mathfrak{C}_N(t)$ cannot scale slower than
$1/N$, independently of whether the  range of the random collisions is
short  or long.   This is  in agreement  with the  fairly good  fit of
\eref{Ctheory}  $\mathcal{C}_N(t)$  for $\delta=5$,  corresponding  to
regime $IV$, shown in the last panel of Fig.~\ref{fig:C-vs-delta}.

In conclusion, our numerical  results suggest that Eq.~(\ref{Ctheory})
yields  a correct  description  of the  autocorrelation  of the  total
energy  current, with  the exception  of  the regime  $III$ for  which
Eq.~(\ref{Ctheory}) is correct up to a time independent multiplicative
constant.  This means  that when  the contribution  of the  noise term
cannot be neglected,  $\mathfrak{C}_N(t)$ contributes to Eq.~(\ref{CN0})
as  a   multiplicative  factor  of  the   deterministic  contribution,
Eq.~(\ref{Ctheory}).  However,  we  can  not  discard  that  the  noise
contribution  $\mathfrak{C}_N(t)$  has a  wider  effect  on the  total
autocorrelation, particularly at longer time scales at which numerical
results converge slowly.

\begin{figure} [!t]
  \centerline{\includegraphics[width=0.85\textwidth]{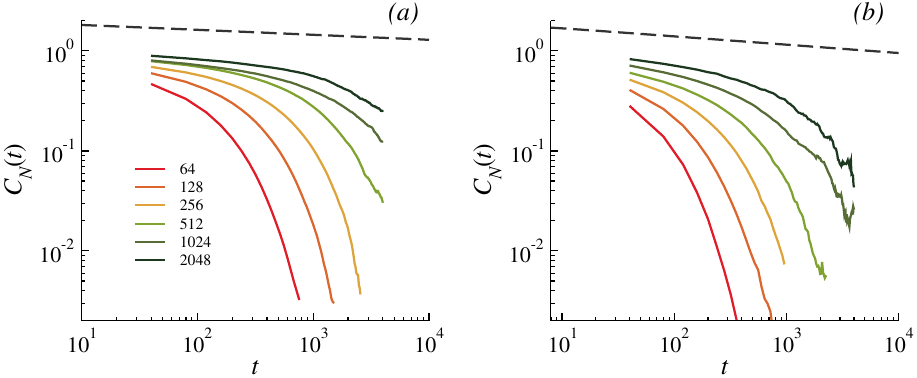}}
  \caption{Autocorrelation function $C_N(t)$ for
    different chain sizes $N$, with $\gamma=0.1$,
    $n_{\mathrm{col}}=4N$, $\delta=2.1$, and
    $\alpha=20$ (panel $a$) and $\alpha=2.2$ (panel $b$). The dashed curves
   correspond to the autocorrelation in the
   thermodynamic limit $C_\infty(t)$.}
 \label{fig:noise-N}
\end{figure}

Moreover, we  cannot discard  either, that  our numerical  results are
subject to additional finite-size effects due to a slow convergence of
the   autocorrelation.    In   Fig.~\ref{fig:noise-N}  we   show   the
autocorrelation function  $C_N(t)$ for different chain  sizes $N$, and
in two different regimes: in panel $a$, $\delta= 2.1$ and $\alpha= 20$
(regime  $II$), and  in panel  $b$,  $\delta= 2.1$  and $\alpha=  2.2$
(regime $III$).  Both regimes exhibit  an extremly slow convergence of
$C_N(t)$  to its  thermodynamic limit  $C_\infty(t)$.

\section{Conclusions}
\label{sec:concl}

In conclusion, we have studied the joint effect of long-range linear 
forces and long-range collisions in a one-dimensional chain
of coupled oscillators.  We relied on the kinetic approach that
 allows to give 
quite easily an approximate decay rate $\mu(k)$ of the normal modes
by Eq.~ (\ref{mu}). Using the small-wavenumber approximations
of $\mu$ and the group velocities $v(k)$, it is than possible to compute analytically
the  autocorrelation of the deterministic heat current.   
Equation (\ref{Ctheory}), along with the expression for the 
exponents (\ref{pq}) is the main result of the work.
From it,  we revealed four possible decay regimes,  
depending of the range exponents $\alpha,\delta$. 
The four regimes are determined by the different scaling 
laws of the two main physical quantities: the group velocity 
and the relaxation rates, as prescribed by Eq. (\ref{C-2}). 
The leading asymptotic decays of $C_N(t)$ are 
summarized in Eq.~(\ref{Csummary}) and 
in in Fig. \ref{fig:phases}.

The asymptotic expansions also reveal that the finite-size 
corrections are very relevant: they decay pretty slowly with
$N$, see Eq. (\ref{leading}), and are definitely sizeable in 
the simulations. We carried out a quantitative check
of the analytical expression against the numerical data.
Besides confirming the validity of the various approximations,   
we believe that  this is a very instructive comparison.  
Indeed, all this should be taken into account when analyzing
the correlation decay in simulations of anharmonic systems.

Assuming that the deterministic contribution $C_N(t)$ dominates 
over the stochastic one,  we may give estimates of the size-dependent
conductivity, following the usual Green-Kubo approach.
Such estimates  hint at a transition 
from anomalous to normal transport (signaled by the dashed line 
in Fig. \ref{fig:phases}) where collision are actually
long-ranged, see Eq.(\ref{kappa}).   This 
may sound counter-intuitive and, 
at least in the region $IV$, it may be explained 
by observing that there waves are overdamped thus hindering
the propagation in favor of energy diffusion.  
A confirmation of this scenario  would require 
a full numerical evaluation of the adequate correlation functions
and/or extensive nonequilibrium simulation, a task that we leave
to future investigations.

\vspace{10pt}
\noindent\textbf{Acknowledgements:} FA thanks SISSA for supporting his stay in Madrid 
through mobility funds for PhD students. 
CMM acknowledges financial support from the Spanish Government grant PID2021-127795NB-I00 (MCIU/AEI/FEDER, UE).
We thank 
Stefano Iubini for  discussions during elaboration of this 
work.

\vspace{10pt}
\noindent\textbf{Data availability:} The datasets generated during and/or 
analysed during the current study are available from the 
corresponding author on reasonable request.

\vspace{1cm}

\section*{References}
\bibliography{long-range.bib}

\appendix

\section{Normal mode decomposition}
\label{sec:fourier}

Consider   a   $2   N$   degrees   of   freedom   Hamiltonian   system
$H(\mathbf{q},\mathbf{p})$, where $\mathbf{q} = (q_1,q_2,\ldots, q_N)$
and  $\mathbf{p}  = (p_1,p_2,\ldots,  p_N)$ are $N$-dimensional
vectors of coordinates and momenta respectively.

The canonical variables can be decomposed in normal modes by taking
\begin{equation} \label{F-exp}
  q_l = \sum_\nu Q_\nu \chi_l^{\nu *} \ , \quad p_l = \sum_\nu P_\nu
  \chi_l^{\nu *} \ ,
\end{equation}
where
\begin{equation} \label{chi}
  \chi^\nu_l = \frac{e^{-ik_\nu l}}{\sqrt{N}} \ , \ \mathrm{and} \ \ \
    k_\nu = \frac{2\pi\nu}{N} \ , \ \ \mathrm{with} \ \ \ \nu = -\frac{N}{2}
    + 1 , \ldots , \frac{N}{2}
\end{equation}
are the Fourier normal modes and the coefficients
\begin{equation} \label{F-coords}
  Q_\nu = \sum_l q_l \chi_l^\nu \ , \quad P_\nu = \sum_l p_l
  \chi_l^\nu \ ,
\end{equation}
the normal mode coordinates satisfying $Q_\nu^* = Q_{-\nu}$,
$P_\nu^* = P_{-\nu}$.
In these coordinates the Hamiltonian \eref{H} becomes
that given in (\ref{F-H}).

Let us perform a further change of variables to
\begin{equation} \label{A}
  \mathbf{A} = i(2\Omega)^{1/2} \mathbf{Q}+(2\Omega)^{-1/2}\mathbf{P} \ ,
\end{equation}
where
\begin{equation} \label{Omega}
  \Omega = \diag(\omega_1,\omega_2,\ldots,\omega_N) \ ,
\end{equation}
is a diagonal matrix of the normal mode frequencies.
In this coordinates, the  Hamiltonian  becomes
\begin{equation} \label{HA}
  H  =  \mathbf{A}^+\Omega \mathbf{A} \ ,
\end{equation}
and more  importantly, the evolution of  vector $\mathbf{A}$ generated
by the Hamiltonian is diagonal
\begin{equation} \label{Afree}
  \mathbf{A}(t+\tau) = e^{i\Omega\tau} \mathbf{A}(t) \ , \quad
  \mathbf{A}^*(t+\tau) = e^{-i\Omega\tau} \mathbf{A}^*(t) \ .
\end{equation}

\section{Relaxation of the energy modes in the Kinetic limit}
\label{sec:kin}

We start by bringing off a further change to  action-angle  variables
$I_{\nu}$,            $\theta_{\nu}$             defined            as
$A_{\nu}=\sqrt{I_{v}}e^{i\theta_{\nu}}$.   The   variation  of   these
variables due to a random collision is
\begin{eqnarray}
    &I_{\nu}^{'} = I_{\nu} \biggr|
      1-\frac{2V_{\nu}e^{-i\theta_{\nu}}}{\sqrt{I_{\nu}}}Z\biggr|^{2}\
      , \label{sto_evol_I}\\
    &\sin \theta_{\nu}^{'} =  \sqrt{\frac{I-{\nu}}{I_{\nu}^{'}}}\sin
      \theta_{\nu}-2\frac{\mathrm{Im}(V_{\nu})}{\sqrt{I_{\nu}^{'}\omega_{\nu}}} \label{sto_evol_theta}
      \ ,
\end{eqnarray}
where as before, the primed variables correspond to their values after
the                           collision,                           and
$Z=\mathrm{Re}(\sum_{\mu}      \sqrt{I_{\mu}\omega_{\mu}}     (V_{\mu}
e^{-i\theta_{\mu}})$.

Plugging in the deterministic evolution of the dynamics \eref{Afree}, we obtain
\begin{eqnarray}
  &I_{\nu}(t+\tau) = I_{\nu}(t)+\Delta I_{\nu}(I,\theta),  \label{sto_I}\\
  &\theta_{\nu}(t+\tau) = \theta_{\nu}(t)+ \omega_{\nu} \tau
    \theta_{\nu}(t)+
    \Delta \theta_{\nu}(I,\theta), \label{sto_theta}
\end{eqnarray}
where $\Delta I$ and $\Delta \theta$ can be read off
\eref{sto_evol_I} and \eref{sto_evol_theta}.  Note that the
action changes only due to the random collisions and, as
expected, are conserved by the deterministic quadratic dynamics.

Now  we consider  the kinetic  limit. Denoting  the mean  time between
succesive random collisions as $\langle\tau\rangle$, the kinetic limit
is      defined      as      the      limit      $N\rightarrow\infty$,
$\langle\tau\rangle\rightarrow\infty$ keeping
\begin{equation} \label{gamma}
  \gamma = \frac{1}{N \langle\tau\rangle} 
\end{equation}
constant.

Noting that in this limit the phases in Eq.~(\ref{sto_theta}) are
randomised in a time scale faster with respect to the evolution of the
actions, it is legitimate to take the average of \eref{sto_I} over a
uniform distribution of the angles, yielding
\begin{equation} \label{sto_evol_Iw}
    \bar{I}_{\nu}^{'}=(1-2|V_{\nu}|^{2})\bar{I}_{\nu}
    +2\frac{|V_{\nu}|^{2}}
    {\omega_{\nu}}\sum_{\mu}\bar{I}_{\mu}\omega_{\mu}|V_{\mu|^{2}} \ .
  \end{equation}
  In terms of the actions, the energy of the normal modes is defined as
  $E_{\nu}=I_{\nu}\omega_{\nu} =\omega_{\nu}|A_{\nu}|^{2}$, and using
  \eref{sto_evol_Iw} we obtain
\begin{equation} \label{sto_E_modes}
  E^{'}_{\nu}=E_{\nu}+\sum_{\mu}K_{\nu \mu}E_{\mu}, \quad
  K^{(n,m)}_{\mu \nu} = -2|V^{(n,m)}_{\nu}|^{2}\delta_{\mu
    \nu}+2|V^{(n,m)}_{\nu}|^{2}|V_{\mu}|^{2},
\end{equation}
where  we have  written the  dependence of  $V$ on  the choice  of the
respective collision  $(n,m)$ explicitly.  From the  definition of $K$
\eref{sto_E_modes} we note that the constant vector $E_{\nu}=E_{eq}$ is
an  eigenvector  of $K$  with  zero  eigenvalue, corresponding  to  the
equilibrium state.

Being $K$ a matrix with random entries, we need to average $E_{\nu}$
over the random collisions occurring between a given time $t$ and
$t+\tau$.  From equation \eref{sto_E_modes} it follows
\begin{equation} \label{sto_E_modes2}
  E_{\nu}(t+\tau)=\sum_{\mu}\biggr(\prod_{\{(n,m)\}}
  (\mathbf{1} + K^{(n,m)} )\biggr)_{\nu\mu} E_{\mu}(t),
\end{equation}
where $\mathbf{1}$ denotes the identity matrix and the product runs
over the collisions between $t$ and $t+T$.  Now we assume that a
single collision alters the energies $E_{\mu}$ only by a small
amount. This is satisfied when the normal modes $\chi^{\nu}$ are given
by the Fourier modes or are otherwise extended, in which case their
components scale as $\sqrt{N}$ due to the normalization condition (see
\eref{chi}). If the eigenmodes are localized this condition might not
hold, but we do not deal with such cases.  In another context, this
approximation is of similar nature to the well-known weak-disorder
expansion, a method used to evaluate the product of random matrices
for small disorder strengths \cite{pikovsky2016}.

Under this assumption \eref{sto_E_modes2} can be linearise yielding
\begin{equation} \label{sto_E_modes2_lin}
    E_{\nu}(t+T)-E_{\nu}(t)=\sum_{\mu}\sum_{\{(n,m)\}}
    K^{(n,m)}_{\nu\mu} E_{\mu}(t) \ .
  \end{equation}
  
  Noting that the sum over the collisions in the right-hand side of
  \eref{sto_E_modes2_lin} is the average of the matrix $K$ over the
  random collisions, $\sum_{\{(n,m)\}} K^{(n,m)} := N\bar{K}$, and
  that the left-hand side of Eq.~\ref{sto_E_modes2_lin} is a time
  derivative $E_{\nu}(t+T)-E_{\nu}(t)\sim \dot{E}_{\nu}(t)/\gamma$,
  \eref{sto_E_modes2_lin} yields the time evolution of the
  normal-mode energies $E_{\nu}$ that can be written in the form of a
  master equation as
\begin{equation} \label{Eevol}
  \dot{E}_{\nu} = \sum_{\mu} (R_{\nu\mu}E_{\mu}-R_{\mu\nu}E_{\nu}) \ ,
\end{equation}
where the matrix $R = \bar{K}/\langle\tau\rangle$ is explicitly
$R_{\mu\nu}=2\gamma N\overline{|{V}_{\nu}|^{2}|V_{\mu}|^{2}}$.

The  evolution  of  the  normal-mode energies  is  determined  by  the
eigenvalues of the linear operator defined by (\ref{Eevol}).  As mentioned  before, the equilibrium
state where the  energy is equipartited over all the  modes leads to a
zero eigenvalue.  Instead, the non-vanishing eigenvalues represent the
relaxation  rates $\mu_\nu$ towards equilibrium. 
For the model studied here, using the definition of $V_{\mu}$,
\eref{V}, we obtain
\begin{equation} \label{Rmatrix}
  R_{\mu\nu} = \gamma \sum_{l>0} W_\alpha(l) \left(-4\sin^{2}
  \frac{k_{\mu}l}{2}\delta_{\mu\nu}+\frac{8}{N}\sin^{2}\frac{k_{\mu}l}{2}\sin^{2}
  \frac{k_{\nu}l}{2} \right) \ ,
\end{equation}
where $W_\alpha(r)=W_\alpha(|n-m|)$ is the  probability of a collision
defined  in \eref{Wprob}.  In  the large  $N$  limit the  off-diagonal
entries of  $R_{\mu\nu}$ are small  with respect to the  diagonal ones
and we  thus neglect  them (this  also means  to neglect  the coupling
between the various energy modes).  In this diagonal approximation the
eigenvalues $\mu_{\nu}$ trivially correspond to  the first term in the
sum of \eref{Rmatrix}, yielding Eq.~(\ref{mu}).

\section{Normal modes at equilibrium}
\label{sec:equil}

Numerically, the initial state of the system is set through the
variables $\mathbf{A}$ and $\mathbf{A}^*$, and to set an equilibrium state at the
initial time $t=0$ means that these variables are distributed according 
to the measure
\begin{equation}
  2\sqrt{\frac{\beta \omega_\nu}{\pi}} \ e^{-\beta \ \omega_\nu |A_\nu|^2} \ \d A_\nu \ \d A_\nu^* \ .
\end{equation}
In polar representation the measure transforms to
\begin{equation}
  2\beta \omega_\nu |A_\nu| e^{-\beta \ \omega_\nu |A_\nu|^2} \ \d
  |A_\nu| \ \frac{1}{2\pi} \ \d \varphi \ .
\end{equation}
This means that to set the system in equilibrium at temperature
$\beta^{-1}$, $A_\nu(0)$ must be chosen 
fixing as a uniform random phase $\varphi$
and a modulus drawn from a Rayleigh
distribution with a scale parameter $1/\sqrt{2\beta\omega_\nu}$.

\section{Correlation function for $\delta=2 $}
\label{sec:delta2}

For $\delta=2$ the energies and velocities of the modes are (in the continuum limit)):
\begin{eqnarray}
    &\omega^{2}(k) \approx \frac{6}{\pi} |k|, \\ 
    &v(k)\approx \frac{\sqrt{6}}{2\sqrt{\pi}} |k|^{-1/2} sign(k),
\end{eqnarray}
while the decay rate is given by:
\begin{equation}
    \mu(k) \approx a_{\alpha}\gamma  |k|^{\alpha-1}.
\end{equation}
The correlation function is then given by:
\begin{eqnarray}
    C_{N}(t)= &\frac{4 (K_{B}T)^{2}}{\pi} \int_{2\pi/N}^{\infty} dk \, v(k)^{2} \, \, e^{-\mu(k)t} \\
    &=\frac{6 (K_{B}T)^{2}}{\pi^{2}} \int_{2\pi/N}^{\infty} \frac{dk}{k}  \, \, e^{-a_{\alpha}\gamma t k^{\alpha-1}} \\
    &=\frac{6 (K_{B}T)^{2}}{\pi^{2}} \left( \frac{-1}{\alpha-1} \right) Ei\left( -\left(\frac{2\pi}{N}\right)^{\alpha-1} a_{\alpha}\gamma t  \right),
\end{eqnarray}
where:
\begin{equation}
    Ei(z)=-\int_{-z}^{\infty} \frac{e^{-t}}{t}dt.
\end{equation}
Expanding in series for small $t/N^{\alpha-1}$ we get:
\begin{eqnarray}
    C_{N}(t)=\frac{6 (K_{B}T)^{2}}{\pi^{2}(\alpha-1)} \left( -\gamma_{E} + \log\left( \left(\frac{N}{2\pi}\right)^{\alpha-1} \frac{1}{a_{\alpha} \gamma t} \right) -S  \right),
\end{eqnarray}
where $\gamma_{E}$ is the Euler-Mascheroni constant, and
\begin{eqnarray}
    S=\sum_{n=1}^{\infty} \frac{1}{n!n} \left(\frac{(-a_{\alpha}\gamma (2\pi)^{\alpha-1} t)}{N^{\alpha-1}}\right)^{n}.
\end{eqnarray}
So for large $N$ we don't  have well defined limit, but a logarithmic divergence like in the short-range collision case \cite{tamaki2020}.

\end{document}